\documentclass[prb,showpacs,preprintnumbers,amsfonts,amssymb,amsmath,floats,twocolumn,aps,superscriptaddress]{revtex4}

\usepackage{graphicx}
\usepackage{dcolumn}
\usepackage{bm}
\usepackage[final,dvips]{epsfig}
\usepackage{color}

\newcommand{\be}{\begin{equation}}
\newcommand{\ee}{\end{equation}}
\newcommand{\ba}{\begin{eqnarray}}
\newcommand{\ea}{\end{eqnarray}}
\newcommand{\ff}[1]{{\bm #1}}

\newcommand{\Tr}{\mbox{Tr}}
\newcommand{\refeq}[1]{Eq.\ (\ref{eq:#1})}
\newcommand{\eqreff}[1]{Eq.\ (\ref{#1})}

\begin{document} 
  
\title{
Variational cluster approach to the Hubbard model:\\
Phase-separation tendency and finite-size effects.
} 

\author{M. Aichhorn}

\affiliation{
Institute for Theoretical Physics, University of
W\"urzburg, Am Hubland, 97074~W\"urzburg, Germany
}

\author{E. Arrigoni}

\affiliation{
Institute of Theoretical Physics and Computational
Physics, Graz University of Technology, Petersgasse 16, 8010
Graz, Austria
}

\author{M. Potthoff}

\affiliation{
Institute for Theoretical Physics, University of
W\"urzburg, Am Hubland, 97074~W\"urzburg, Germany
}
\affiliation{
Institute for Theoretical Physics, University of
Leipzig, vor dem Hospitaltore 1, 04103 Leipzig, Germany
}

\author{W. Hanke}

\affiliation{
Institute for Theoretical Physics, University of
W\"urzburg, Am Hubland, 97074~W\"urzburg, Germany
}

\begin{abstract}
Using the variational cluster approach (VCA), we study the transition from the 
antiferromagnetic to the superconducting phase of the two-dimensional 
Hubbard model at zero temperature.
Our calculations are based on a new method to evaluate the VCA grand potential
which employs a modified Lanczos algorithm and avoids integrations 
over the real or imaginary frequency axis.
Thereby, very accurate results are possible for cluster sizes not accessible 
to full diagonalization.
This is important for an improved treatment of short-range correlations, 
including correlations between Cooper pairs in particular.
We apply this improved method in order to investigate the cluster-size dependence of the phase-separation tendency
that has been proposed recently on the basis of calculations for smaller 
clusters. While the energy barrier 
associated with phase separation  rapidly decreases  with
increasing cluster size for both hole and electron doping,
the  extension of the phase-separation region behaves
differently  in the two cases.
More specifically, our results suggest that  phase separation
remains persistent in the hole-doped and disappears in the
electron-doped case. We also study 
the evolution of the single-particle spectrum as a function of doping
and point out the relevance of our results for experimental findings in electron
and hole-doped materials.

\end{abstract} 
 

\maketitle

\section{Introduction}

Since the discovery of high-temperature superconductivity in copper-based
transition-metal oxides, 
a tremendous effort has been devoted to establish a convincing theory
that covers the general aspects of their unusual and fascinating physics.
The attempts are complicated by the fact that strong electron correlations 
play a key role in the physics of the cuprates.
A central question in this context concerns the emergence of small energy scales, 
much smaller
than the bare (Coulomb) interactions between the electrons, which govern the 
existence and the competition of different phases at low temperatures.
This can be studied by considering prototypical lattice models of strongly
correlated electrons.
Some agreement has been achieved that the relevant physics of the cuprate
high-temperature superconductors is covered by the two-dimensional one-band 
Hubbard model,\cite{an.87}
\begin{equation}\label{eq:2Dhub}
  H = \sum_{i,j,\sigma} t_{ij} c^{\dagger}_{i\sigma} c^{\phantom{\dagger}}_{j\sigma} +
  U \sum_{i} n_{i\uparrow} n_{i\downarrow}-\mu\sum_in_i \: .
\end{equation}
Here $t_{ij}$ denote the hopping matrix elements, $n_{i\uparrow}$ is the density 
at site $i$ with spin ``$\uparrow$'', $n_i=n_{i\uparrow}+n_{i\downarrow}$, $\mu$ the 
chemical potential, and $U$ the local Coulomb repulsion.

In recent years there has been substantial progress in the
understanding of the ground-state properties of 
the Hubbard model due to the development of quantum-cluster theories,\cite{ma.ja.05_rev}
such as cluster extensions of the dynamical mean-field theory (DMFT), i.e.\ 
the dynamical cluster approximation \cite{he.ta.98} and the
cellular DMFT \cite{ko.sa.01,li.ka.00}, or the variational cluster
approach (VCA).\cite{po.ai.03,da.ai} 
These cluster calculations confirm the fact 
that the ground state away from half-filling has a non-vanishing
superconducting order
parameter\cite{se.la.05,ai.ar.05,ai.ar.06,ma.ja.05,ca.ko.06} 
with a pairing interaction
of predominantly d-wave character.\cite{ma.ja.sc.06} 
Recent VCA calculations\cite{se.la.05,ai.ar.05,ai.ar.06} suggest that
at low electron and hole doping the two-dimensional Hubbard model is in a 
symmetry-broken mixed AF+SC state where both the antiferromagnetic (AF)
and the superconducting (SC) order 
parameters are finite. 
This is consistent with recent cellular DMFT calculations.\cite{ca.ko.06} 
When going to higher dopings, the system
displays a tendency to phase separate into an AF+SC phase
at lower doping and a pure SC phase at higher doping.

The VCA accesses the physics of a lattice model in the thermodynamic 
limit by optimizing trial self-energies generated by a reference system.
The above-mentioned VCA calculations are based on a reference system 
consisting of small ($2\times 2$) isolated clusters tiling the infinite
lattice. 
This generates trial self-energies which are very short ranged spatially. 
Hence, there is the obvious question for the robustness of the
results as a function of the size of the individual clusters.
Up to now, it was not possible to consider larger cluster sizes and,
at the same time, reach a sufficient accuracy to resolve the tiny
energy scale driving phase separation, especially in the electron-doped 
case.
The reason is that an accurate evaluation of the VCA grand potential has
required a {\em full} diagonalization of the cluster Hamiltonian 
(see Ref.~\onlinecite{ai.ar.06} for details) which has severely restricted 
the available cluster sizes.

The purpose of this paper is to present a new method for the 
evaluation of the VCA grand potential based on the Lanczos method which 
leads to sufficiently accurate results even for larger clusters where 
full diagonalization is no longer possible.
Using this method we investigate the competing phases in the two-dimensional 
Hubbard model at zero temperature for clusters up to 10 sites. 
This implies a substantial qualitative step forward as short-range correlations 
between different Cooper pairs can be included -- opposed to calculations based
on $2\times2$ clusters.

\section{Variational cluster approach}

The variational cluster approach is one of the possible approximation schemes
that can be constructed within the self-energy-functional theory (SFT).
\cite{pott.epj}
The SFT provides a variational scheme to use dynamical information from an 
exactly solvable ``reference system'' (for example an isolated cluster) to 
approximate the physics of a system in the thermodynamic limit. 
For a system with Hamiltonian $H=H_0(\ff t)+H_1(\ff U)$ and one-particle and 
interaction parameters $\ff t$ and $\ff U$, the grand potential is written as 
a functional of the self-energy $\ff \Sigma$ as
\begin{equation}
\Omega[\ff \Sigma] = F[\ff \Sigma] + \mbox{Tr} \ln 
\left(\ff G_{0}^{-1} - \ff \Sigma\right)^{-1} \: ,
\end{equation}
with the stationary property 
$\delta \Omega[\ff \Sigma_{\rm phys}] = 0$ for the 
physical self-energy. Here, $\ff G_{0}=(\omega + \mu - \ff t)^{-1}$ 
is the free Green's function of the original model
in the thermodynamic limit at frequency $\omega$, and
$F[\ff \Sigma]$ is the Legendre transform of the universal 
Luttinger-Ward functional. Due to its universality it is the same as 
the functional for a ``simpler'' problem with the same interaction but
a modified one-particle part $\ff t'$. The stationary solutions are obtained 
{\em within} the subspace of self-energies 
$\ff \Sigma = \ff \Sigma(\ff t')$ of the simpler 
problem that is spanned by varying $\ff t'$. 
This restriction constitutes the approximation. 
Details of the approach are described in Refs.\ \onlinecite{pott.epj,pott.ass}.

The VCA \cite{po.ai.03} is generated within the SFT by choosing 
as a reference system a set of isolated clusters which tile up the original 
infinite lattice.
By construction, the VCA correctly incorporates correlation effects in the 
electron self-energy up to the length scale given by the cluster size. 
Beyond this scale it acts like a mean-field approximation.
One of the main advantages of the VCA as compared to the simpler cluster
perturbation theory \cite{se.pe.00} consists 
in its ability to describe (normal and off-diagonal) long-range order by 
including suitably chosen fictitious symmetry-breaking Weiss fields 
in the set of variational parameters.
Microscopically coexisting phases can be obtained using several Weiss
fields.
The method links in a consistent way the static 
thermodynamics with the frequency-dependent one-particle excitation spectra 
(photoemission).
Details of the approach have been described elsewhere. 
\cite{da.ai,ai.ar.05,ai.ar.06}

The VCA grand potential to be calculated in practice reads as 
\begin{equation}\label{eq:ocalc}
  \Omega = \Omega^\prime + {\rm Tr}\,{\rm ln}
  \left(\ff G_0^{-1} - \ff \Sigma\right)^{-1}-
       {\rm Tr}\,{\rm ln}\left(\ff G^\prime\right) \: .
\end{equation}
Here, $\ff G_0$ is the free Green's function of the model given by \refeq{2Dhub},
$\Omega^\prime$, $\ff \Sigma$, and $\ff G^\prime$ are the grand potential, 
the self-energy and the Green's function of the cluster reference system
which depend on the one-particle parameters $\ff t'$.
In the present study we 
consider clusters with $L_c=4$, 8, and 10 sites to search for the
stationary solution characterized by the condition $\partial \Omega / \partial 
\ff t^\prime = 0$.~\cite{maxcl}
This stationary point provides a good approximation to the exact
solution for the system in the thermodynamical 
limit if the self-energy is sufficiently ``short ranged'',
i.e.\ sufficiently localized within the cluster. 

\begin{figure}
  \centering
  \includegraphics[width=0.75\columnwidth]{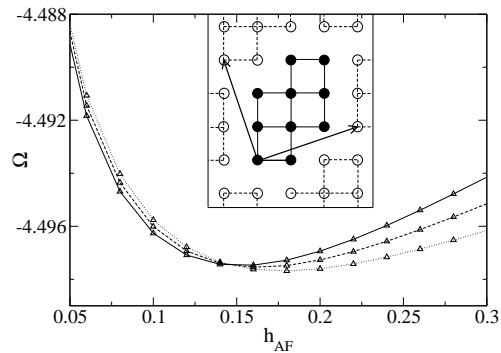}
  \caption{\label{fig1}%
    The SFT grand potential $\Omega$ of the half-filled ($n=1$)
    Hubbard model with nearest-neighbor hopping $t=-1$ and $U=8t$
    as a function of the variational parameter $h_{\rm AF}$ (staggered
    magnetic Weiss field). Reference system (inset): $L_c=10$ clusters.
    We compare results obtained by integration over real frequencies 
    with Lorentzian broadenings $\eta=0.1$ 
    (dotted lines) and $\eta=0.05$ (dashed lines), as well as for the
    ``$Q$-matrix'' evaluation (see text, solid lines). 
    $S_L=100$ Lanczos iteration steps have been performed. 
  }
\end{figure} 
    
As discussed in Refs.~\onlinecite{ai.ar.05,ai.ar.06}, it is important to evaluate 
$\Omega$ with high accuracy in order to resolve the relevant
energy scales of the competing phases, especially in the
electron-doped case.
Here, we present a method in which this evaluation can be done without a 
numerical integration over frequencies (note that frequency integration is
implicit in the $\Tr \ln \cdots$ terms in \refeq{ocalc}).
We start from the expression derived in
Ref.\ \onlinecite{pott.epj} which expresses the $\Tr \ln \cdots$ terms 
as a sum over the single-particle excitation energies:
\begin{align}
\Tr \ln\, \left(\ff G_0^{-1} - \ff \Sigma\right)^{-1} &\stackrel{\phantom{T=0}}{=}
- \sum_m T \ln (1+{\rm e}^{-\beta \omega_m}) - R\nonumber\\
&\stackrel{T=0}{=}\sum_m \omega_m \Theta(-\omega_m) -R
\label{eq:poles}
\end{align}
and 
\begin{align}
\Tr \ln\, \ff G^\prime &\stackrel{\phantom{T=0}}{=}
- \sum_m T \ln (1+{\rm e}^{-\beta \omega^\prime_m}) - R \nonumber\\
&\stackrel{T=0}{=}\sum_m \omega^\prime_m \Theta(-\omega^\prime_m) -R\; .
\end{align}
Here $\Theta(\omega)$ is the Heaviside step function, $\beta=1/T$ the 
inverse temperature.
$\omega'_m$ are the one-particle excitation energies of the reference 
system, i.e.\ the poles of $\ff G^\prime$, and $\omega_m$ are the poles 
of the VCA Green's function $(\ff G_0^{-1} - \ff \Sigma)^{-1}$.
$R$ represents a contribution due to the poles of the self-energy (see Ref.\ 
\onlinecite{pott.epj}) which cancels out in Eq.\ (\ref{eq:ocalc}) and can 
thus be ignored.
The excitation energies $\omega'_m = E_r - E_s$ of the reference
system (i.e. of the cluster) can be readily obtained with the help of 
the Lanczos algorithm from the eigenenergies $E_r$ of the reference system.
Here, we introduce the notation $m=(r,s)$, to indicate an excitation
between two states $s$ and $r$.
The major difficulty consists in finding the poles $\omega_m$ of the
VCA Green's function. 

This can be done in the following way:
Consider the Lehmann representation\cite{fe.wa.71} of 
$\ff G^\prime$ which can be cast into the form\cite{za.ed.02}
\begin{equation}
  G^\prime_{\alpha\beta}(\omega)=
  \sum_{m}
  Q_{\alpha m} \frac{1}{\omega - \omega^\prime_m} Q^\dagger_{m\beta} \; ,
\end{equation}
where $\alpha$ refers to the one-particle orbitals of the cluster (typically
$\alpha=({\rm site}\:i,{\rm spin}\:\sigma)$ but it can also include
an orbital index).
The ``$Q$-matrix'' is defined as:
\begin{align}
  Q_{\alpha m} &\stackrel{\phantom{T=0}}{=}
  \langle r | c_\alpha | s \rangle 
  \sqrt{\frac{\exp(-\beta E_r) + \exp(-\beta E_{s})}{Z^\prime}}\nonumber\\
  &\stackrel{T=0}{=}\delta_{r,0}\langle 0 |c_\alpha |s\rangle + 
  \delta_{s,0}\langle r|c_\alpha|0\rangle\; . 
\label{qam}
\end{align}
The spectral weight (residue) of $G'_{\alpha\beta}(\omega)$ at a pole 
$\omega=\omega^\prime_m$ is given by $Q_{\alpha m}Q^\dagger_{m\beta}$. 
$Z^\prime=\sum_r e^{-\beta E_r}$ is the grand-canonical partition function
at finite temperature, and $|0\rangle$ denotes 
the (grand-canonical) ground state of the reference system.
Introducing the diagonal matrix $g_{mn}(\omega)=\delta_{mn} / (\omega
- \omega^\prime_m)$, 
we have: 
\begin{equation}
\ff G^\prime(\omega) = \ff Q \ff g(\omega) \ff Q^\dagger \; .
\label{gpq}
\end{equation}

Defining $\ff V = \ff t - \ff t^\prime$, which in case of the VCA typically
includes the inter-cluster hopping terms, 
the ``subtraction'' of the fictitious Weiss fields, 
as well as shifts
of the one-particle energies (see below), \cite{ai.ar.06} the VCA 
expression for the lattice Green's function can be written as:
\begin{equation}
  \ff G \equiv \frac{1}{\ff G_0^{-1} - \ff \Sigma} 
  =
  \frac{1}{(\ff G^\prime)^{-1} - \ff V} \: .
\end{equation}
This expression can be transformed with the help of the $\ff
Q$-matrix Eq.~(\ref{qam}) and Eq.~(\ref{gpq}):
\begin{align}
  \ff G &= \frac{1}{(\ff Q \ff g \ff Q^\dagger)^{-1} - \ff V} \nonumber \\
  &=\ff Q \ff g \ff Q^\dagger + \ff Q \ff g \ff Q^\dagger \cdot \ff V \cdot
  \ff Q \ff g \ff Q^\dagger  + \cdots \nonumber \\
  &= \ff Q \left(\ff g + \ff g \cdot \ff Q^\dagger \ff V \ff Q \cdot \ff g + \cdots
  \right) \ff Q^\dagger \nonumber \\
  &= \ff Q \frac{1}{\ff g^{-1} - \ff Q^\dagger \ff V \ff Q} \ff Q^\dagger \: .
  \label{eq:matrix}
\end{align}
Note that $\ff Q$ is not a square matrix  and that $\ff Q \ff Q^\dagger = \ff 1 
\neq \ff Q^\dagger \ff Q$.
Since $\ff g^{-1} = \omega - \ff \Lambda$ with $\Lambda_{mn}= \delta_{mn} \omega^\prime_m$, 
the poles of $\ff G$ are now simply given by the eigenvalues
of the (frequency independent) 
matrix $\ff M= \ff \Lambda + \ff Q^\dagger \ff V \ff Q$ and can be easily found by 
diagonalization.
The dimension of $\ff M$  
is given by the number of poles of $\ff G^\prime$ with non-vanishing spectral 
weight.\cite{somezero} Hence, the above scheme 
requires the knowledge of all
excited 
states of the reference system.
In Refs.~\onlinecite{ai.ar.05,ai.ar.06}, these states have been obtained by a 
{\em full} diagonalization of a rather small ($2\times2$) cluster.

For larger clusters, where a full diagonalization is not possible, 
the Lanczos algorithm should, in principle, provide precisely the required
poles and matrix elements \eqreff{qam}. In practice, however, there are
some difficulties, as we discuss below.
Within the Lanczos method the matrix elements 
$G_{\alpha\beta}(\omega) =\langle\langle c^{\phantom{\dagger}}_\alpha;
c^\dagger_\beta\rangle\rangle_\omega$ of a 
cluster Green's function at $T=0$ are determined in $2L_c$ separate Lanczos 
procedures.\cite{comment} In each procedure, one takes as a Lanczos
initial vector one element of the sets
$\left\{ c^{\phantom{\dagger}}_{1,\sigma}|0\rangle,
\cdots,c^{\phantom{\dagger}}_{L_c,\sigma}|0\rangle\right\},\left\{c^\dagger_{1,\sigma}|0\rangle,\cdots,
c^\dagger_{L_c,\sigma}|0\rangle\right\}$ where
$|0\rangle$ is the cluster ground state. 
In principle the poles should be the same for all matrix elements of the 
Green's function.
In practice, however, the poles obtained by the $2L_c$ runs
are slightly different from each other
due to the limited numerical accuracy of the Lanczos method.
Therefore, this kind of Lanczos algorithm is not suited for the 
``$Q$-matrix'' evaluation of the grand potential described above, since 
merging all matrix elements of $\ff G'$ into the compact form \eqreff{gpq}
would results in a too large matrix $\ff M$ that cannot be diagonalized.

Fortunately, the problem can be overcome by means of the so-called 
{\em band Lanczos} method.\cite{bandlanczos} 
The difference with respect to the
standard algorithm is that the sets of initial vectors given above are used 
{\em simultaneously} within {\em one single} Lanczos run. This yields the same
set of poles for all index pairs $(\alpha,\beta)$ as well as the corresponding 
weights. 
The dimension of the matrix $\ff M$ is given by the number of iteration steps
$2S_L$ in the Lanczos procedure. 
In this case, one only needs two Lanczos procedures instead of $2L_c$. 
Using this method, one introduces an error due to the limited set ($2S_L$) 
of the excited states in the reference system that are kept in the Lanczos
calculation.
Generally, however, this error is extremely small since excitations with
large weight result from states which converge very fast with increasing 
$S_L$. These excitations with large weight, on the other hand, are just 
those which are dominant in \refeq{ocalc} compared to excitations with 
small weight.\cite{pott.epj}

We have checked the accuracy of our method by considering the symmetry-broken 
antiferromagnetic phase of the Hubbard model at half-filling (Fig.~\ref{fig1}). 
One can clearly see that the $Q$-matrix evaluation perfectly gives the 
extrapolation of results obtained by numerical frequency integration with 
finite but small Lorentzian broadenings $\eta$. 
We also verified that the results converge very fast with $S_L$, i.e., 
typically $S_L\approx 100$ is fully sufficient.
Last but not least, this improved method substantially reduces the
computational time.
For example, a factor of approximately 15 is gained for the $L_c=10$ cluster.

\section{Results for Competing Phases}

\begin{figure}
  \centering
  \includegraphics[width=0.67\columnwidth]{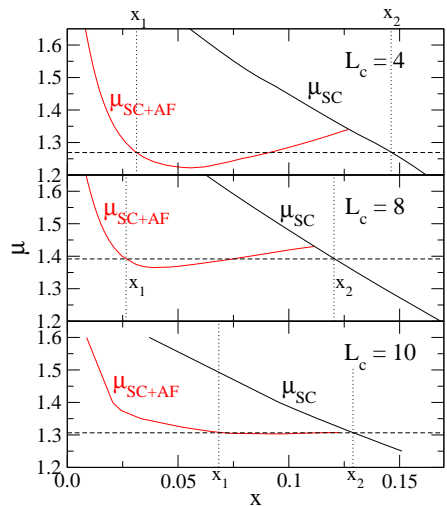}
  \caption{\label{fig2}%
    (Color online) Chemical potential $\mu$ as function of hole doping $x$. Results
    for $L_c=4$ ($2\times 2$), $L_c=8$ ($4\times 2$), and $L_c=10$ clusters. The 
    horizontal dashed lines mark the critical $\mu_c$, and the
    vertical dotted lines mark the boundaries $x_1$ and $x_2$ of the phase 
    separation region in between.
  }
\end{figure}
\begin{figure}
  \centering
  \includegraphics[width=0.67\columnwidth]{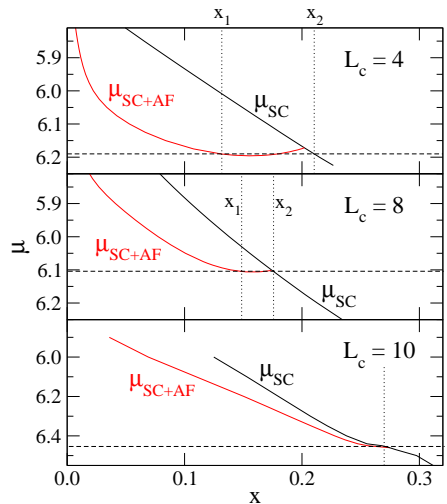}
  \caption{\label{fig3}%
    (Color online) Same as Fig.\ \ref{fig2} but for electron doping. 
    Note that there is no phase 
    separation for $L_c=10$; the dotted line marks the quantum critical point. 
  }
\end{figure}
\begin{figure}
  \centering
  \includegraphics[width=0.7\columnwidth]{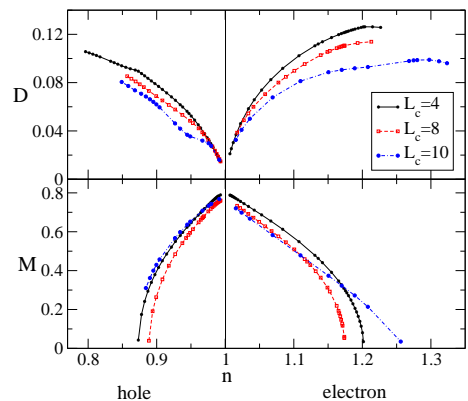}
  \caption{\label{fig4}%
    (Color online) Magnetization $M$ and d-wave order parameter $D$ as functions 
    of hole and electron doping for $L_c=4$, 8, and 10.
  }
\end{figure}

On the basis of this improved evaluation, we investigate the finite-size behavior 
of the phase-separation tendency observed for small clusters in Ref.~\onlinecite{ai.ar.05}.
As variational parameters we use the Weiss fields 
$h_{\rm AF}$ and $h_{\rm SC}$ to allow for antiferromagnetic (AF) and d-wave 
superconducting (SC) orders, respectively,\cite{da.ai,ai.ar.05}
as well as an overall shift $\varepsilon$ of the one-particle energies in the cluster
to ensure a consistent treatment of the particle density.\cite{ai.ar.06} 
Fig.~\ref{fig2} shows our results for the two-dimensional Hubbard model with $U=8t$ 
and next-nearest-neighbor hopping $t^\prime = -0.3t$ for the case of hole doping. 
The calculations have been performed for $L_c=4$ (top), $L_c=8$ (middle), and 
$L_c=10$ clusters (bottom). 

\begin{table}
  \centering
  \caption{\label{tab:jumps_orderpar}%
    Discontinuities $\Delta x$, $\Delta M$, and $\Delta D$ across the PS region for hole- 
    and electron doping.}
  \begin{tabular}{|c|c|c|c|}
    \hline
    {\bf h-dop.} &  $\Delta x$ & $\Delta M$ & $\Delta D$\\
    \hline
    $L_c=4$  & 0.115 & 0.717 & 0.055 \\
    $L_c=8$  & 0.094 & 0.699 & 0.043 \\
    $L_c=10$ & 0.056 & 0.568 & 0.032 \\
    \hline
  \end{tabular}
  \begin{tabular}{|c|c|c|c|}
    \hline
    {\bf e-dop.} &  $\Delta x$ & $\Delta M$ & $\Delta D$\\
    \hline
    $L_c=4$  & 0.079 & 0.476 & 0.016 \\
    $L_c=8$  & 0.020 & 0.316 & 0.004 \\
    $L_c=10$ & 0.000 & 0.000 & 0.000 \\
    \hline
    \end{tabular}
\end{table}

We consider the case of hole-doping first. One can immediately see that the 
phase-separation tendency, found for the $L_c=4$ cluster, 
weakens progressively when increasing the cluster size.
In particular, the corresponding ``energy scale''
$\Delta\mu=\mu^\ast-\mu_c$ diminishes very rapidly with increasing $L_c$
($\Delta\mu=0.050$ for $L_c=4$, $\Delta\mu=0.027$ for $L_c=8$,
and  $\Delta\mu=0.003$ for $L_c=10$), and appears to vanish in the
$L_c\to\infty$ limit. 
Here, $\mu^\ast$ is
the point where the slope of $\mu(x)$ changes sign and $\mu_c$ the chemical potential 
at the transition point. 
However, this fact does not necessarily imply the absence of  macroscopic phase
separation in the exact ground state of the model under study. As a
matter of fact,
the exact function 
$\mu(x)$ {\em must} 
have a non-positive slope.
Therefore, in the phase-separated
case $\mu(x)$ becomes a straight line between the two boundaries of the
phase-separation region $x_1$ and $x_2$.\cite{he.ma.99}
Whether the exact ground state supports phase separation can be
derived from the finite-size scaling of the doping discontinuity
$\Delta x \equiv x_2-x_1$, see Tab.~\ref{tab:jumps_orderpar}.
Unfortunately, no regular finite-size behavior can be inferred
from Fig.~\ref{fig2} and Tab.~\ref{tab:jumps_orderpar} for hole doping, 
probably due to the fact that the clusters are
still too small. 
Opposed to the clear trend visible for the electron-doped case (see
Tab.~\ref{tab:jumps_orderpar}), there is a much weaker $L_c$ dependence
of the discontinuities $\Delta x$, $\Delta M$, and $\Delta D$, which we rather 
interpret as being irregular. However, our results do not exclude microscopic
phase separation to persist for $L_c\to\infty$.
The inclusion of long-range Coulomb interaction would then be necessary in
order to ``frustrate'' the phase separation occurring in the plain
Hubbard model and produce microscopic
inhomogeneous phases, such as stripes.\cite{kive,em.ki.93,lo.em.94}
We stress that 
at this point only qualitative estimates for $L_c\to\infty$ rather than a 
convincing finite-size scaling are possible.
For a discussion on these issues see, e.g., 
Refs.~\onlinecite{he.ma.99,wh.sc.03,be.ca.00}.

The situation is quite different in the electron-doped case. 
Here, not only the phase-separation energy $\Delta
\mu$, but also  the doping discontinuity $\Delta x$ 
appears to vanish for $L_c\to\infty$.
In fact,  $\Delta \mu$ is
already an order of magnitude smaller than for hole doping in the $L_c=4$ 
cluster.~\cite{ai.ar.05}
In addition, already for $L_c=10$ the transition from the AF+SC to the pure SC
phase has become
continuous at least within numerical accuracy.
In this case, the weak phase separation observed at the mean-field
level for small clusters was simply a signal of a 
{\em tendency} of the system to produce {\em
microscopically} inhomogeneous phases (such as stripes), as
conjectured in Ref.~\onlinecite{ai.ar.05}.
The fact that the corresponding energy scale is already very small for
a small cluster
could explain why
there is no clear sign of stripes in electron-doped materials and could
possibly be related to the much smaller pseudogap energy scale, as
discussed in Ref.~\onlinecite{ai.ar.05,ai.ar.06}. 

\begin{figure}
  \centering
  \includegraphics[width=0.6\columnwidth]{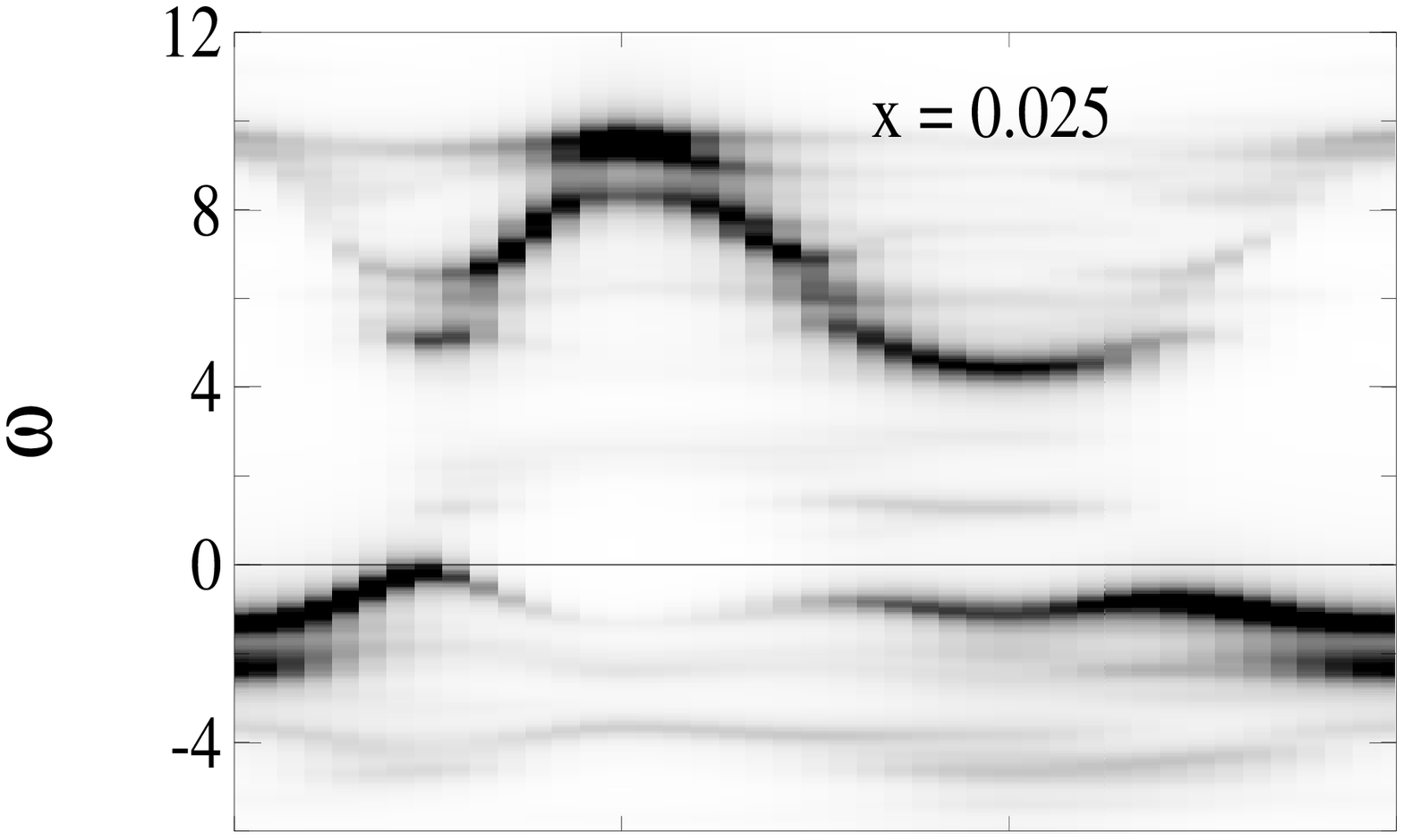}\\
  \includegraphics[width=0.6\columnwidth]{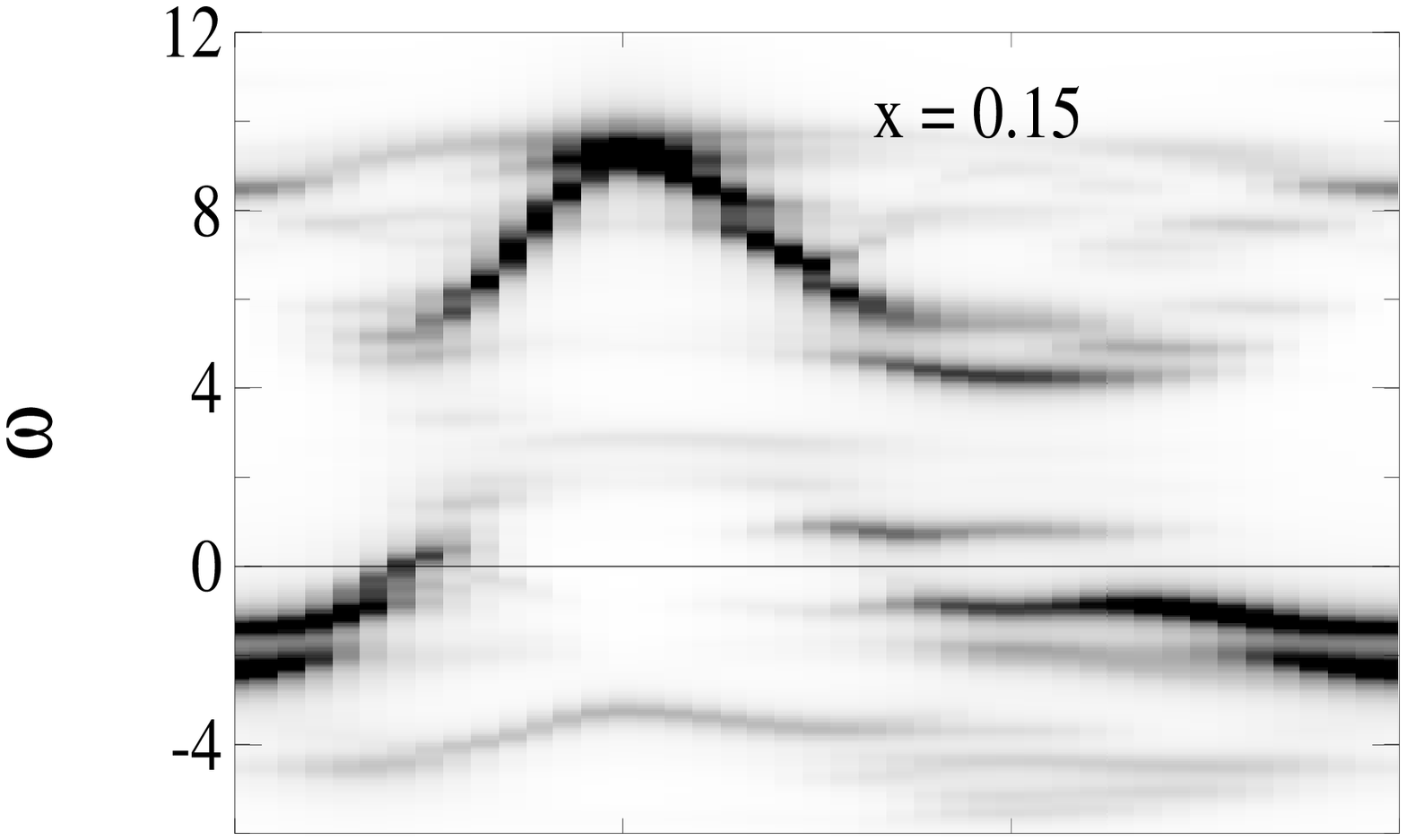}\\
  \includegraphics[width=0.6\columnwidth]{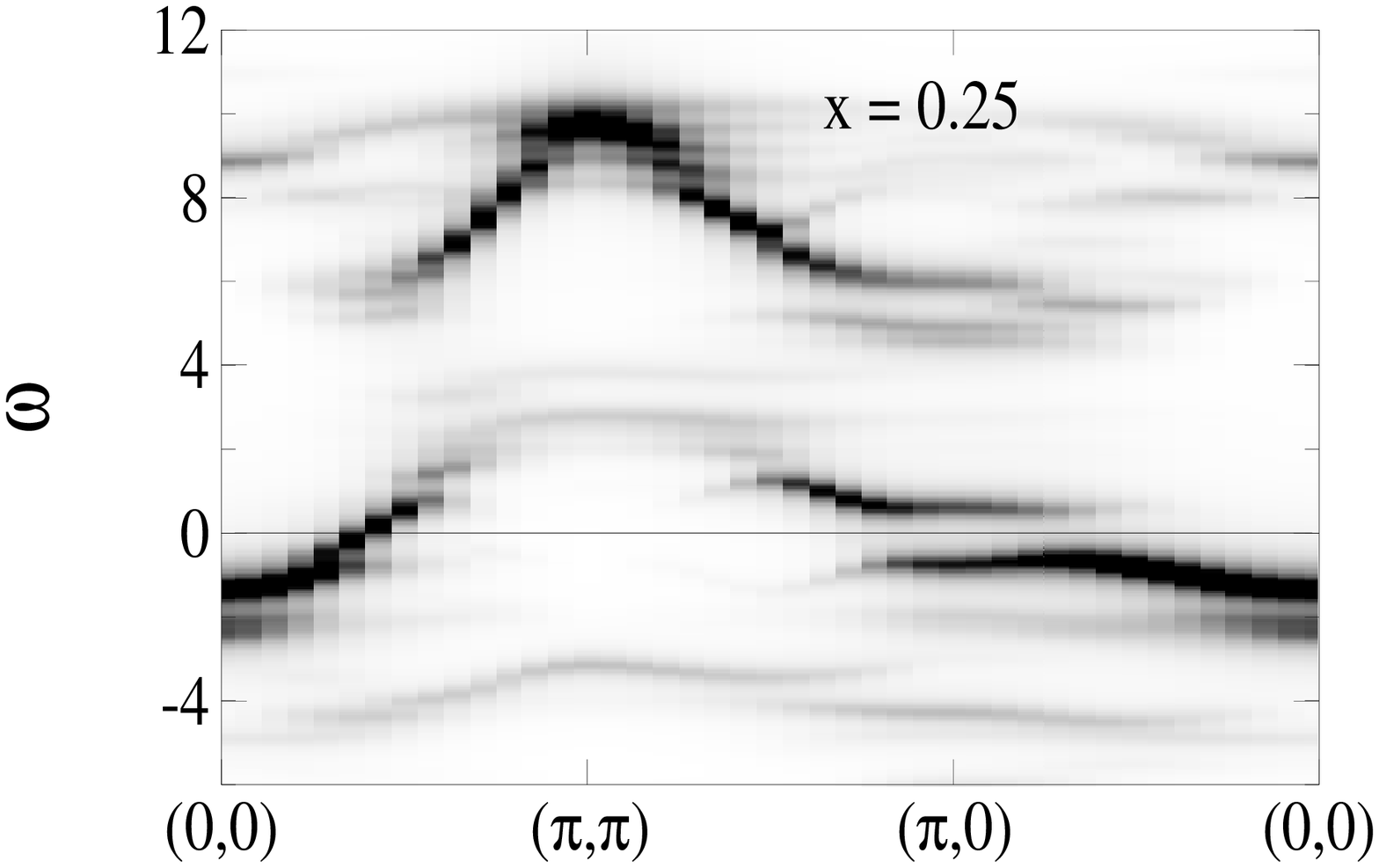}
  \caption{\label{fig5}%
    Evolution of the spectral function upon hole doping ($L_c=8$). Top panel:
    doping $x=0.025$, mixed SC+AF phase. Middle panel: $x=0.15$, SC phase. 
    Bottom panel: $x=0.25$, SC phase. 
    A Lorentzian broadening of $\eta=0.2t$ has been used to display the results.
  }
\end{figure}

Contrary to the phase-separation energy, the AF and SC order
parameters $M$ and $D$ plotted in Fig.~\ref{fig4}
only display a rather weak cluster-size dependence.
This shows that already a small $2\times2$ cluster describes the static
ground-state quantities with a rather good accuracy, except for cases 
close to a phase transition.
Finite-size effects are more pronounced for $D$ because the 
SC order parameter is a non-local quantity which converges slower with increasing
cluster size as compared to $M$.
Nevertheless, from our results we can argue that for both, hole and electron doping,
at least substantial SC fluctuations remain in the thermodynamic limit even in the 
AF phase. 
A more precise finite-size scaling to identify as to whether one really has long-range 
SC order is not possible since that would involve larger constant cluster shapes 
($2\times 2$, $4\times 4$,...) which are not accessible by the present VCA. 

The comparison of Fig.~\ref{fig4} with previous calculations\cite{se.la.05} shows 
that the inclusion of the energy shift $\varepsilon$ as a variational
parameter provides results which depend only weakly on the cluster size $L_c$.
For example, we find a mixed AF+SC phase for small doping
for all cluster sizes considered. This can be understood by the fact that 
the inclusion of additional variational parameters
``optimize'' the ground state of the reference system
towards the exact solution of the infinite lattice.

\begin{figure}
  \centering
  \includegraphics[width=0.6\columnwidth]{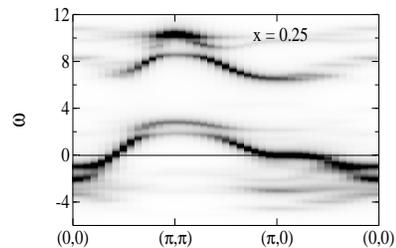}
  \caption{\label{fig6}%
    Spectral function for $x=0.25$ ($L_c=8$), SC phase, as in Fig.\ \ref{fig5} 
    but for a different stationary point with a particle density $n'=0.75...$ in the cluster.
  }
\end{figure}

Fig.~\ref{fig5} shows the evolution of the single-particle
spectral function upon hole doping calculated with the $L_c=8$ 
reference system.
In the upper plot for doping $x=0.025$, the system is still in the mixed AF+SC phase. 
This is the reason for the ``back-turning'' of the quasi-particle-like band around
$(\pi/2,\pi/2)$ although the chemical potential already ``touches''
the band at this wavevector. 
For higher doping (middle and lower panel), 
we clearly see a transition to a dispersion crossing
the chemical potential in the nodal direction,
in agreement with angle-resolved photoemission experiments.
The low-energy coherent quasi-particle band with a width of the order of a few times $J$ has
been replaced by a band of width of a few times $t$.
The qualitative trend
is well known from QMC calculations.\cite{qmc} This represents a clear improvement as compared
to our previous results for $L_c=4$ clusters, where the dispersion in the nodal
direction showed ``back-turning'' signals also for higher dopings.

In spite of the improvement for the nodal direction, the d-wave SC gap appears to be too large 
for the slightly overdoped case $x=0.25$. Furthermore, for higher dopings one expects a decrease 
of the weight of the upper Hubbard band which is much stronger than visible in the $x=0.25$ spectrum. 
The reason for these shortcomings is probably a too strong admixture of the half-filled cluster 
ground state: In the absence of superconductivity and for not too high doping, the particle 
density of the reference system (the isolated cluster) is $n'=1$. In our case, deviations from 
cluster half-filling are introduced due to a non-vanishing SC Weiss field only. For $L_c=8$ we 
find $n'=0.92$ at $x=0.25$ which is still close to half-filling. 

A physically better description of the spectral density at higher dopings can only be achieved 
when (in the absence of superconductivity) starting from a cluster ground state with $n'<1$. This 
yields a corresponding SFT grand potential which has to be compared with the SFT grand potential 
for the $n'=1$ stationary point. Note that for a vanishing SC Weiss field, the VCA cannot give a 
grand potential that is continuous in the entire doping range. This is an artifact of the VCA which 
levels off and eventually becomes irrelevant in the large-cluster limit. 

In fact, there is a second stationary point for $x=0.250$ with a particle density in the cluster 
$n'=0.755$. Due to the non-vanishing SC Weiss field, this is close but not equal to the commensurate 
cluster filling $n'=0.75$. This stationary point, however, exhibits an SFT grand potential which is 
{\em higher} as compared to that of the $n'=0.92$ solution and, consequently, should be disregarded. 
It is nevertheless interesting to discuss the spectral density of this (metastable) state which is 
shown in Fig.\ \ref{fig6}. As could have been expected, the SC gap is much smaller (and actually not 
visible on the scale of the figure) and, as compared to the corresponding spectrum in Fig.\ \ref{fig5}, 
the weight of the upper Hubbard band is clearly reduced. Moreover, signatures of magnetic order are no longer
visible in this spectrum.

\section{Conclusions}

We have developed a new method to evaluate the VCA grand potential
which avoids numerical integrations over real or Matsubara
frequencies, even for large clusters, for which a complete
diagonalization is not feasible.
This provides a sufficient accuracy to study the cluster-size
dependence of the phase-separation tendency obtained in previous works. 
The results of the present paper suggest that in the hole-doped case phase separation
observed for small clusters persists for $L_c\to\infty$, i.e. in the exact ground state,
while it eventually disappears in case of electron doping.
This would explain why there is no clear sign  of stripes in electron-doped materials, 
and could possibly be related to the much smaller (or even absent) pseudogap 
energy scale with respect to hole-doped materials.\cite{ai.ar.05,al.kr.03}

\section*{Acknowledgements}

We thank S. A. Kivelson for enlightening comments, in particular for
pointing out the correct interpretation of the finite-size scaling of
$\Delta \mu$.
This work was supported by the Deutsche Forschungsgemeinschaft within the 
Forschergruppe FOR538 and, partially, by the Austrian Science Fund (FWF 
projects P18505-N16 and P18551-N16), as well as by the KONWIHR 
supercomputing network in Bavaria.


\begin{thebibliography}{99}
\bibitem{maxcl}
Notice that the maximum cluster size we can reach is smaller than in
conventional Lanczos calculations. This is due to the fact that (i) we do not
conserve particle number within the cluster in the SC phase, (ii) we
have open boundary conditions within the cluster, and (iii) the
cluster exact diagonalisation has to be reiterated many times before
convergence in the three variational parameters (see below) has been achieved.
\bibitem{somezero} Occasionally, one of the poles of $\ff M$ may have
  a vanishing residue in Eq.~(\ref{eq:matrix}). 
  In this case, one can show that $\ff G'$ has a corresponding pole at the same frequency
  with a vanishing residue so that the 
  contribution to Eq.(\ref{eq:ocalc}) from this pole cancels out exactly.
\end{thebibliography}
\end{document}